\font\msbm=msbm9  scaled \magstep 1
\def\kappa{\hbox{\msbm\char"7B}}         

\def\title #1{\centerline{\rm #1}}
\def\author #1: #2;{\line{} \centerline{#1}\smallskip\centerline{#2}}
\def\abstract #1{\line{} \centerline{ABSTRACT} \line{} #1}

\newcount\refno \refno=1
\documentstyle[12pt]{article}
\begin{document}
\centerline{}
\centerline{}

\moveright 20pt
\vbox{

\centerline{\quad \bf {First and second order phase transitions and magnetic
hysteresis}}

\centerline{\bf { for a superconducting plate }}

\centerline{}

\centerline{ G. F. Zharkov}
\centerline{}
\centerline{ {\it \ \ P.N. Lebedev Physical Institute, Russian Academy of
Sciences, Moscow, 119991, Russia} }
\centerline{} \centerline{}    }

{\hsize =15.5truecm
\moveright 15pt
\vbox{
\it{
The self-consistent solutions of a nonlinear Ginzburg--Landau equations, which
describe the behavior of a superconducting plate of thickness $2D$ in a
magnetic field $H$ parallel to its surface (provided that there are no vortices
inside the plate), are studied. We distinguish two types of superconductors
according to the behavior of their magnetization $M(H)$ in an increasing field.
The magnetization can vanish either by a first order phase transition (type-I
superconductors), or by a second order (type-II). The boundary $S_{\rm I-II}$,
which separates two regions (I and II) on the plane of variables ($D,\kappa$),
is found. The boundary $\zeta(D,\kappa)$ of the region, where the hysteresis
in a decreasing field is possible (for superconductors of both type), is also
calculated. The metastable $d$-states, which are responsible for the hysteresis
in type-II superconductors, are described. The region of parameters ($D,\kappa$)
for type-I superconductors is found, where the supercooled normal metal (before
passing to a superconducting Meissner state) goes over into a metastable
precursor state ($p$-). In the limit $\kappa\to 1/\sqrt{2}$ and $D\gg\lambda$
(where $\lambda$ is the London penetration depth) the self-consistent
$p$-solution coincides with the analytic solution, found from the degenerate
Bogomolnyi equations. The critical fields $H_1,H_2,H_p,H_r$ for type-I and
type-II superconducting plates are also found.

\par } } }

{\hsize=16.5truecm

\centerline{}
\centerline{\bf 1. Introduction}
\centerline{}

The macroscopic Ginzburg--Landau theory of superconductivity (GL) [1] is widely
used to describe the superconductors behavior in the external magnetic field.
In particular, in [1] the structure of the intermediate region in the
vicinity of superconducting ($s$-) and normal ($n$-) semi-infinite metallic
phases, brought into contact in magnetic field, was studied. It was found that
the free energy $\sigma_{ns}$ of the interface between $n$- and $s$-phases
vanishes, when GL-parameter $\kappa$ equals to $\kappa_0=1/\sqrt{2}$. On
this basis all superconductors are usually divided in two groops [1]: the
superconductors with $\sigma_{ns}>0$ ($\kappa<\kappa_0$) belong to first
groop, the superconductors with $\sigma_{ns}<0$ belong to second groop. (The
value $\sigma_{ns}<0$ indicates that $s$-phase is unstable in respect to the
appearance of vortices in a bulk superconductor [2].)

However, one can use different criterion to divide superconductors in two
groops, namely, according to the shape of the magnetization curve, using the
formula $\overline{B}=H+4\pi M$, where $\overline{B}$ is the mean field value
in the specimen, $H$ is the external field, $M$ is the magnetization. Such
criterion was used in [3] for a cylinder of radius $R$ (assuming no vortices
inside the cylinder). On the base of the self-consistent solutions of
GL-equations the critical parameter $\kappa_c(R)$, dividing two types of the
dependence $M(H)$ in the icreasing field, was found. It was shown that for
$\kappa<\kappa_c$ (in type-I superconductors) the cylinder magnetization
vanishes in a jump (by a first order phase transition to $n$-state). For
$\kappa>\kappa_c$ (in type-II superconductors) the magnetization vanishes
gradually, by a second order phase transition. (The difference in critical
parameters $\kappa_0$ and $\kappa_c(R)$ is due to the different geometries and
different criteria for sorting the solutions, assumed in [1,2] and [3].)\, It
was also shown in [3] that the cylinder behavior in a magnetic field is by far
nontrivial: there exist several states (the solution branches), the transitions
between these states, the hysteresis phenomena, etc. It was noted that the
properties of the vortex-free state should be taken into account in
interpreting the experiments with sufficiently thin (mesoscopic) samples.

In this paper we consider an infinite plate of thickness $2D$ in a field $H$
parallel to its surface, assuming no vortices inside the plate. (Such
vortex-free state was previously studied numerically in [4,5], but not in
sufficient detail.)\, We show that even in the simplest case of a vortex-free
plate the solutions depend on parameters $(D,\kappa,H)$ in a very complicated
way, analogously to the case of a cylinder geometry [3,6,7]. Thus, there exists
a critical value $\kappa_c(D)\ne\kappa_0$, which defines two types of the
magnetization behavior, $M(D,\kappa,H)$. In type-I superconductors (with
$\kappa<\kappa_c$) the magnetization of the plate vanishes (if the field
increases) in a jump (by a first order phase transition into $n$-state in
some field $H_1$). In type-II superconductors (with $\kappa>\kappa_c$) the
magnetization vanishes gradually (by a second order phase transition into
$n$-state in a field $H_2$). If the thickness is sufficiently small,
$D\sim\lambda$ ($\lambda$ is the London penetration depth), the
superconducting state is destroyed by a second order phase transition at
arbitrary $\kappa$. If $\kappa=\kappa_c(D)$, the first and second order phase
transitions become indistinguishable. We found also, that type-II
superconducting plate, which is in the vortex-free Meissner state
(with the order parameter $\psi\sim 1$), may pass (if the field increases,
FI-regime) into a special (also vortex-free) "edge-suppressed" $e$-state [5].
The order parameter $\psi$ in $e$-state is strongly suppressed in some layer
near the plate surface, so the magnetic field $B$ in this layer is not screened
and practically equals to the external field $H$. In type-I superconductors
such $e$-state does not form. (The possibility of transitions into $e$-state
means that alongside with the usual vortex mechanism [2] there exists the
addidional "edge" mechanism of the field penetration into a mesoscopic sample.)

The hysteresis $d$-states, which appear in type-II superconductors if the field
decreases (FD-regime), and the so-called "precursor" states ($p$-) in type-I
superconductors, which describe the onset of $s$-state (in the field $H_p$)
from the supercooled $n$-state, are also studied. We show, that in the limit
$D\gg\lambda$ and $\kappa\to\kappa_0=1/\sqrt{2}$ the hysteresis (metastable)
$p$-solution coincides with the solution of the degenerate Bogomolnyi
equations [8] and can be described analytically [9].

The paper is divided into several Sections. In Sec. 2 the basic equations and
boundary conditions are written, and necessary notations introduced; in Sec. 3
the critical lines $S_{\rm I-II}$, $\zeta$ and $\pi$, which exist on a plane
of parameters $(D,\kappa)$, are calculated and explained; in Sec. 4 the
superconductor behavior in the vicinity of the critical lines $S_{\rm I-II}$
and $\zeta$ is described; in Sec. 5 the examples of the space profiles of
different solutions are given; in Sec. 6 the metastable solutions, which
exist in the region of $p$-states and on its $\pi$-boundary are studied; in
Sec. 7 the connection of the self-consistent solutions of GL-equations with
the solutions of the degenerate Bogomolnyi equations is discussed; in Sec. 8
the critical fields $H_1$, $H_2$, $H_p$ and $H_r$, which characterize the
behavior of type-I and type-II superconducting plates, are found; in Sec. 9
the results are shortly discussed.

\centerline{}
\centerline{\bf 2. Equations}
\centerline{}

In the case of a plate the system of GL-equations [1] can be written in the
following dimensionless form:
$$
{ {d^2a} \over {dx^2} } - \psi^2 a=0,                          \eqno(1)
$$
$${ d^2 \psi \over dx^2 }+\kappa^2 (\psi-\psi^3)-a^2\psi=0.     \eqno(2)
$$
Here, instead of the dimensioned potential $A(x')$, field $B(x')$, current
$j_s(x')$ and the co-ordinate $x'$, the dimensionless quantities $a(x)$, $b(x)$
and $j(x)$ are introduced, where
$$
A={ {\phi_0} \over {2\pi\lambda} } a,\ \,
B={ {\phi_0} \over {2\pi\lambda^2} }b,\ \,
b={da \over dx},\ \,
j(x)=j_s\Big/ { {c\phi_0} \over {8\pi^2\lambda^3} }= -\psi^2 a
                                                              \eqno(3)
$$
($x=x'/\lambda$ is the dimensionless co-ordinate,
$-D_\lambda \le x\le D_\lambda$, $D_\lambda=D/\lambda$, $\lambda=\kappa\xi$,
$\xi$ is the coherence length, $\kappa$ is the GL-parameter, $\phi_0=hc/2e$ is
the flux quantum).

We will consider the vortex-free state. In this case the order parameter
$\psi(x)$ is an even real function, and the potential $a(x)$ (and the current
$j(x)$) are the odd functions of co-ordinate: $a(x)=-a(-x)$, i.e. $a(0)=0$.
Consequently, the boundary conditions to Eq. (1) can be written in the form
(assuming $0\le x \le D_\lambda$):
$$
a\big|_{x=0} = 0,\quad
\left. { {da}\over{dx} }\right|_{x=D_\lambda}=h_\lambda,       \eqno(4)
$$
where $h_\lambda=H/H_\lambda$, $H_\lambda=\phi_0/(2\pi\lambda^2)$.

As to the Eq. (2), we will take the usual boundary condition at the external
surface [1]: $d\psi / dx|_{x =D_\lambda}=0$. The order parameter at the center
is maximal, thus, the boundary conditions to Eq.(2) are:
$$
 \left. { {d\psi}\over dx }\right|_{x = 0} = 0, \quad
 \left. { {d\psi}\over {dx} } \right|_{x = D_\lambda} =0.        \eqno(5)
$$

The magnetic moment (or magnetization) of the plate, related to the unity
volume, is
$$
{M\over V}={1\over V}\int  { B-H \over 4\pi }dv =
{ B_{av}-H \over 4\pi }, \quad
B_{av}={1\over V}\int B({\bf r})dv={1\over S}\Phi(D_\lambda)={1\over D}A(D),
$$
where $B_{av}$ is the mean field value inside the superconductor, $S$ is the
plate cross-section in the $(x,y)$-plane. In normalization (3), denoting
$\overline{b}=B_{av}/H_\lambda$, $M_\lambda=M/H_\lambda$, one finds:
$$
4\pi M_\lambda=\overline{b}-h_\lambda,\quad
\overline{b}={1\over D_\lambda}a_\lambda, \quad
a_\lambda=a(D_\lambda), \quad D_\lambda={D\over\lambda}.          \eqno(6)
$$

The difference of Gibbs free energies of the system in superconducting and
normal states, $\Delta G=G_s-G_n$, can be expressed through its magnetic
moment:
$$
\Delta G={\cal F}_{s0}-{1\over 2}MH,\quad
{\cal F}_{s0}={ H_c^2 \over 8\pi }\int \left[
\psi^4 -2\psi^2+\xi^2\left( {d\psi \over dx'} \right)^2 \right] dv, \eqno(7)
$$
where ${\cal F}_{s0}$ corresponds to the superconductor condensation energy,
$H_c=\phi_0/(2\pi\sqrt{2}\lambda\xi)$ is thermodinamical critical field. Using
(3), one finds from (7) the normalized expression
$$
\Delta g=\Delta G\Big/ \left( { H_c^2 \over 8\pi } V \right)=
g_0-{8\pi M_\lambda h_\lambda \over \kappa^2},\quad
g_0={1\over D_\lambda} \int_0^{D_\lambda} dx \left[ \psi^4-2\psi^2+
{1\over\kappa^2} \left({d\psi \over dx} \right)^2 \right].     \eqno(8)
$$
The expressions (6)--(8) are used below.

[Note, that the lengthes $\xi$ and $\lambda=\kappa\xi$ enter the GL-equations
on equal footing, each one may be chosen as the unit of length; for measuring
the field the unit may be either $H_\lambda=\phi_0/(2\pi\lambda^2)$, or
$H_\xi=\phi_0/(2\pi\xi^2)$ ($H_\xi=\kappa^2 H_\lambda$). In presenting the
numeric results various variants of normalization will be used.]

The solutions were found by using the iteration procedure, described in [10].
In the beginning of iterations the trial function was taken as
$\psi(x)\approx 1$ in FI-regime, and $\psi(x)\approx 0.01$ in FD-regime.
The results do not depend on the choice of the concrete numeric algorithm.

If $\psi\ll 1$ ($B(x)\approx H$), the system of equations (1), (2) reduces to
a single $\kappa$-independent linear equation
$$
d^2\psi/dx^2+(1-h_\xi^2 x^2)\psi(x)=0                     \eqno(9)
$$
($-D_\xi\le x\le D_\xi$, $D_\xi=D/\xi$, $h_\xi=H/H_\xi$), with
$\psi(x)=e^{-y^2}w(y)$, $y=\sqrt{2h_\xi}x$, where the Weber function $w(y)$
satisfies the confluent hypergeometric equation $w''-yw'-aw(y)=0$,
$a={1\over 2}(1-1/h_\xi)$, and can be expressed in a form [11],(2.44):
$$
w(y)=C_1w_1(y)+C_2w_2(y),                                 \eqno(10)
$$
$$
w_1(y)=1+\sum_{\nu=1}^\infty { a(a+2)\cdots(a+2\nu-2) \over (2\nu)!} y^{2\nu},
$$
$$
w_2(y)=y\left[ 1+\sum_{\nu=1}^\infty
{ (a+1)(a+3)\cdots(a+2\nu-1) \over (2\nu+1)! } y^{2\nu} \right].
$$
Eq. (9) and the boundary conditions (5) allow to find the critical field of the
second order phase transition, when $\psi(x)\to 0$ [12,13]. This critical field
does not depend on $\kappa$. However, in the case of first order phase
transitions the condition $\psi(x)\ll 1$ is not fulfilled (see below), so in
a general case to find the critical fields it is necessary to solve the full
system of nonlinear equations (1), (2), (5).

\centerline{}
\centerline{\bf 3. The state diagram on the plane $(D_\lambda,\kappa)$}
\centerline{}

The solutions of Eqs. (1)--(5) depend on the co-ordinate $x$ and three
parameters $(D_\lambda,\kappa,h_\lambda)$. In Fig. 1 the state diagram on the
plane of parameters $(D_\lambda,\kappa)$ is presented. In each point of this
plane there exist a set of self-consistent solutions for the order parameter
$\psi(x;h_\lambda)$ and the field $b(x;h_\lambda)$. If the representation point
$(D_\lambda,\kappa)$ shifts, the corresponding state changes. In particular,
the magnitude of the order parameter in the origin of co-ordinates changes,
$\psi_0=\psi(0;h_\lambda)$, and also the mean field value
$\overline{b}=h_\lambda+4\pi M_\lambda$. Such information permits to judge
how the solution depends on the external field. It is convenient to imagine a
peep-hole pierced in arbitrary point $(D_\lambda,\kappa)$, what allows to
see the dependence of $\psi_0$ (and of the magnetization,
$-4\pi M_\lambda=h_\lambda-\overline{b}$) on the field $h_\lambda$ in this
point. Studying such dependences, one can find on the plane of parameters
$(D_\lambda,\kappa)$ three critical lines ($\pi$, $S_{\rm I-II}$ and $\zeta$),
which divide this plane into four regions (I$_a$, I$_b$, II$_a$ and II$_b$)
with its own characteristic behavior of $\psi_0$ (and $-4\pi M_\lambda$) on
the field $h_\lambda$. The meaning of these regions is clarified in Fig. 2.

Fig. 2(a) shows (schematically) the dependence $\psi_0(h_\lambda)$ in region
I$_a$. Evidently, for small $h_\lambda$ there exists a stable superconducting
state with $\psi(x)\approx 1$ ($\psi_0\approx 1$), in which a weak external
field is almost competely screend and does not penetrate into the bulk of a
superconductor (the Meissner state, M-). With the field increasing (FI-regime)
the order parameter is gradually suppressed, but when the field $h_\lambda$
exceeds the critical value $h_1$  M-solution becomes unstable (small
perturbations grow) and passes in a jump ($\delta_1$) into normal state
($n$-). For $h_\lambda>h_1$ there is only one stable $n$-solution with
$\psi\equiv 0$.

Now, searching for the solution in FD-regime with small values of the trial
function $\psi(x)\ll 1$, one finds that $n$-state remains stable at
$h_\lambda<h_1$ down to the restoration field $h_r$ in Fig. 2(a). In the
field interval $\Delta_n=h_1-h_r$ two stable states (M- and $n$-) exist, but a
supercooled $n$-state is metastable, because M-state (with $\psi\approx 1$)
has smaller free energy (due to the negative condensation energy $g_0$ in (8)).
At $h_\lambda=h_r$ $n$-state looses stability (small perturbations grow) with
a consequent jump ($\delta_r$) into M-state. For $h_\lambda<h_r$ there is only
one stable M-solution.

The analogous picture is present in Fig. 2(e), where initial linear grows of
the magnetization ($-4\pi M_\lambda$) corresponds to the Meissner state
($\overline{b}\approx 0$); the transition from M- to $n$-state is acompanied
by a first order jump ($\delta_1$); there is a hysteresis loop due to the
presence of a supercooled $n$-state, and a first order jump ($\delta_r$)
from $n$- to M-state.

The characteristic behavior $\psi_0(h_\lambda)$ in region I$_b$ (Fig. 1) is
depicted schematically in Fig.2(b). The picture here is analogous to Fig. 2(a):
in FI-regime there is a jump ($\delta_1$) from M- to $n$-state; however, a
supercooled (metastable) $n$-state in the field $h_p$ passes into a special
(also metastable) precursor ($p$-) state, which looses stability in the field
$h_r$ with a first order jump ($\delta_r$) to M-state.

The behavior of the magnetization in this region (Fig. 2(f)) is also
characterized by a jump ($\delta_1$); by the presence of a supercooled
$n$-state in the field interval $\Delta_n=h_1-h_p$; by the presence of
$p$-state (which exists in the field interval $\Delta_p=h_p-h_r$), and by a
jump ($\delta_r$) from $p$- to M-state in the field $h_r$. The total width of
the hysteresis loop is $\Delta_{pn}=\Delta_p+\Delta_n=h_1-h_r$. The width of
the interval $\Delta_p$ diminishes in the vicinity to the critical $\pi$-line
(Fig. 1), and $\Delta_p=0$ on $\pi$-line. Thus, $p$-states exist only in
region I$_b$.

In region II$_a$ M-state in FI-regime (see Fig. 2(c)) becomes unstable in
the field $h_1$ and transforms in a jump ($\delta_1$) into a new stable
"edge-suppressed" state ($e$-). In this state the order parameter is strongly
suppressed in some layer near the plate boundary, so the magnetic field
penetrates this layer practically without screening (see below Fig. 6). If
$h_\lambda$ is further increased, the superconducting $e$-state is destroyed
gradually ($\psi\to 0$), with final transition to $n$-state in the field $h_2$.

The plate magnetization (Fig. 2(g)) is also characterized by the presence of
$e$-tail and by a second order phase transition to $n$-state in the field $h_2$.
If now the field is decreased, $n$-state becomes absolutely unstable in the
field $h_2$, where $e$-state appears again and for fields $h_\lambda<h_1$ it
passes smoothly into a metastable "depressed" ($d$-) state (see Fig. 6). Such
$d$-state (i.e., a supercooled $e$-state) is a characteristic feature of region
II$_a$, it exists in the field interval $\Delta_d=h_1-h_r$ alongside with
M-state and is responsible for a magnetization hysteresis (Fig. 2(g)). In
region II$_a$ a supercooled $n$-state is absolutely unstable, in difference to
region I$_b$ (Fig. 2(b,f)) where $n$-state is metastable in the field interval
$\Delta_n=h_1-h_p$.

If the thickness $D_\lambda$ decreases, remaining in region II$_a$
($\kappa={\rm const}$), the jump amplitudes $\delta_r$, $\delta_1$ (and the
interval $\Delta_d=h_1-h_r$, Fig. 2(c,g)) decrease also, and they vanish
($\delta_r=\delta_1=\Delta_d=0$) on $\zeta$-line (Fig. 1). Below $\zeta$-line
(in region II$_b$) the hysteresis is impossible (Fig. 2(d,h)) and $s$-state
passes into $n$-state by a second order phase transition.

The solutions behavior in the vicinity of critical lines $\pi$, $S_{\rm I-II}$
and $\zeta$ will be studied in more details below, here we note the following.
As is clear from Figs. 2(e-h), a superconductor in the increasing field passes
to $n$-state either by a first order jump, or there is a tail of $e$-states on
the magnetization curve and $n$-state appears by a second order phase
transition. Acordingly, we will distinguish type-I and type-II superconductors.
The boundary between two types of superconductors is represented by the curve
$S_{\rm I-II}$ (Fig.1). This boundary (or, equivalently, the critical value
$\kappa_c(D_\lambda)$) depends on the plate thickness, what does not coincide
with the simple criterion $\kappa_0=1/\sqrt{2}$ [1], used usually for
dividing superconductors into two groops. This disagreement (as in the case
of a cylinder [3]) is caused by several reasons.

First, in [1] the case is considered of an infinite superconductor, while we
consider a superconducting plate of finite thickness. Second, our
superconductor boarders a vacuum, while in [1] a case is considered of two
contacting semi-infinite $s$- and $n$-metals. Third, we divide two types of
superconductors according to the shape of their magnetization curves, while in
[1] the division is made using different criterion: according to the sign of
the surface free-energy $\sigma(\kappa)$ of the interface (with $\sigma=0$ at
$\kappa=1/\sqrt{2}$ [1]). Thus, the mentioned disagreement is due to different
settings of the problem.

\centerline{}
\centerline{\bf 4. $S_{\rm I-II}$- and $\zeta$-boundaries; $G$-point }
\centerline{}

In this section the results of a self-consistent calculations for magnetization
$-4\pi M_\lambda(h_\lambda)$ and $\psi_0(h_\lambda)$ are given, in a case of a
superconducting plate with the parameters $D_\lambda, \kappa$ laying near
$S_{\rm I-II}$-boundary in Fig. 1.

Fig. 3 represents the case $D_\lambda=7$ for three values of $\kappa$. In
Figs. 3(a,b) ($\kappa=0.9$, a peephole is in region I$_b$) there are: M-state,
a jump $\delta_1$ from M- to $n$-state (in the field $h_1$, FI-regime); a
supercooled $n$-state ($\Delta_n$); a metastable $p$-state; a jump $\delta_r$
from $p$- to M-state (in the field $h_r$, FD-regime).

In Figs. 3(c,d) ($\kappa=0.9193$, a peephole is on $S_{\rm I-II}$-boundary)
a supercooled $n$-state has vanished already, but the tail of $e$-states did
not yet appear; this special $p$-state (or marginal $\mu$-state) attains
maximum amplitude in FD-regime at the restoration field $h_r$ (with a jump to
M-state).

In Figs. 3(e,f) ($\kappa=0.95$, a peephole is in region II$_a$) there is a jump
($\delta_1$) from M- to $e$-state, a supercooled $n$-state is absent, however,
a metastable $d$-state appears which ends in a jump ($\delta_r$) from $d$- to
M-state in the field $h_r$.

There is a hysteresis loop on all the curves in Fig. 3; solid lines correspond
to FI-regime, dotted lines correspond to FD-regime.

Fig. 4 illustrates what happens for smaller plate thickness ($D_\lambda=2$).
The value $\kappa=0.93$ belongs to region I$_b$ in Fig. 1. The value
$\kappa=0.953$ corresponds to the boundary $S_{\rm I-II}$. The value
$\kappa=0.98$ corresponds to region II$_a$ (the field interval $\Delta_d$,
where a hysteresis $d$-state exists, diminishes in moving closer to $\zeta$-boundary).
The value $\kappa=1.03$ lies on $\zeta$-boundary, here the jumps vanish
($\delta_1=\delta_r=\Delta_d=0$) and the curves become hysteresis-less, having
a vertical tangent in point $i_0$. For $\kappa=1.05$ (region II$_b$) there is
no hysteresis, but the inflexion point $i$ with finite derivative remains on
the curves. If $D_\lambda$ diminishes further, the inflexion point lowers and
the curves become monotonous, without inflexion.

It is interesting also to trace what happens in the region of $G$-point in
Fig. 1. The critical lines $S_{\rm I-II}$, $\zeta$ and $\pi$ merge in this
point ($\kappa_G\approx 0.915$, $D_G\approx 1.51$) and for $D_\lambda<D_G$
there is a single critical curve. Above it (in region I$_a$) the destruction
(and nucleation) of superconductivity is acompanied by jumps $\delta_1$ (and
$\delta_r$). Below it (in region II$_b$) there is a smooth second order phase
transition. Thus, for sufficiently small thicknesses all type-I superconductors
(with $\kappa<\kappa_G$) become, in fact, type-II superconductors.

To the same conclusion (basing on different considerations) arrived Ginzburg
[14], who noticed that type-I superconductors (with $\kappa\ll 1$) behave in
magnetic field as type-II superconductors. Therefore, $G$-point may be refered
to as the Ginzburg point. [Apart $G$-point there exist the so called
tricritical Landau points ($L$-) [15], where the distinction vanishes between
the critical fields, which correspond to the supercooled, equilibrium and
superheated states of a superconductor [16] (i.e., where the hysteresis
vanishes). The hysteresys-less critical line $\zeta$ in Fig. 1 consists, in
fact, of $L$-points.]

Fig. 5(a) shows the dependence of $\psi_0$ and $-4\pi M_\lambda$ on the field
$h_\lambda$ in $G$-point. These curves end by a second order phase transition
to $n$-state, having vertical tangent at the transition point $h_2$. Fig. 5b
illustates the dependence $\psi_0(h_\lambda)$ in the vicinity of $G$-point
($D_G=1.51$): curve {\it 1} corresponds to $\kappa=0.8$ (region I$_a$ in
Fig. 1); curve {\it 2} corresponds to $\kappa_G=0.915$ ($G$-point); curve
{\it 3} corresponds to $\kappa=1.0$ (region II$_b$). In region I$_a$ (curve
{\it 1}) there exists a supercooled $n$-state ($\Delta_n$), there are first
order jumps ($\delta_1$ and $\delta_r$) and the hysteresis loop. On curve
{\it 2} the derivative $d\psi_0/dh_\lambda=\infty$ at the second order phase
transition field $h_2$. The hysteresis-less curve {\it 3} with finite
derivative at the transition point $h_2$ corresponds to region II$_b$.

\centerline{}
\centerline{\bf 5. Examples of the co-ordinate dependences }
\centerline{}

Figs. 1--5 illustrate the solutions behavior on parameters
$(\kappa, D_\lambda, h_\lambda)$. In Fig. 6 the self-consistent solutions
$\psi(x)$ and $b(x)$ are depicted as functions of the reduced co-ordinate
$x/D_\lambda$, when the representation point crosses the plane of parameters
($\kappa,D_\lambda$) in Fig. 1 along the lines $D_\lambda=7$ and $D_\lambda=2$.

Fig. 6a shows the space profile of the order parameter $\psi(x)$ for
$D_\lambda=7$ and several vallues of $\kappa$. Curve M$_e$ ($\kappa=1.1$,
region II$_a$ in Fig. 1) is the Meissner state in the field $h_1=0.9742$,
which preceeds the jump to $e$-state ($h_\lambda=0.9743$). The order parameter
of $e$-curve is suppressed near the plate boundary. When the field $h_\lambda$
is further increased, the amplitude of $e$-state tends to zero and vanishes
finally at $h_2=1.2114$. Curve $p$ ($\kappa=0.8$, region I$_b$) is a precursor
state in the field $h_r=0.5676$, which preceeds the jump to the Meissner state
M$_p$ ($h_\lambda=0.5675$). Curve $\mu$ ($\kappa=0.9193$, the peephole is on
$S_{\rm I-II}$-boundary) corresponds to the marginal $p$-state, which attains
the maximal amplitude ($\psi_r=0.988$) before transforming into M$_\mu$-state
($\psi_0=0.9999$) at $h_r=0.6520$. The depressed $d$-state forms in FD-regime
from $e$-state (which exists at $h_\lambda=0.9743$, $\kappa=1.1$) by gradual
transformation of $e$-solution profile and ends (at $h_r=0.7797$) in a jump
to the Meissner M$_d$-state ($h_\lambda=0.7796$).

The corresponding profiles of $b(x)$ are depicted in Fig. 6(b).

The solutions $\psi(x)$ and $b(x)$ for a plate of smaller thickness
($D_\lambda=2$) are depicted in Figs. 6(c,d). Shown are: $e$-solution at the
point of transition from M$_e$- to $e$-state ($\kappa=0.97$, region II$_a$);
$d$-solution at the point of transition from $d$- to M$_d$- ($\kappa=0.97$,
region II$_a$); $\mu$-solution in the field of transition from $\mu$- to
M$_\mu$- ($\kappa=0.953$, the boundary $S_{\rm I-II}$); $p$-solution in the
field of transition from $p$- to M$_p$- ($\kappa=0.93$, region I$_b$).
The solution $i_0$ is also shown, which lies on $\zeta$-boundary
($\kappa=1.030$, $h_\lambda=1.076$) with $M'=\infty$ at the inflexion point,
and the solution $i$ ($\kappa=1.05$, region II$_b$, $h_\lambda=1.1$) with
finite value $M'$ at the inflexion point.

\centerline{}
\centerline{\bf 6. Region of $p$-states, $\pi$-boundary}
\centerline{}

In this Section the precursor solutions $\psi(x)$ and $b(x)$ are shown at the
points ($\kappa, D_\lambda$) belonging to region I$_b$ of Fig. 1.

The space profiles $\psi(x)$ and $b(x)$ for a plate with $D_\lambda=10$ and
various $\kappa$ are depicted in Figs. 7(a,b). The value $\kappa=0.922$
corresponds to $S_{\rm I-II}$-boundary (at $D_\lambda=10$). At this point the
jump from M- to $n$-state happens in the field $h_1=0.8511$ (FI-regime, see
Fig. 3(c)). In FD-regime a supercooled $n$-state does not exist (it is
absolutely unstable), but a superconducting $p$-solution appears, having the
amplitude $\psi_0$ which grows with the field diminishing and reaches the
maximum value in the field $h_r=0.6529$. Such hysteresis $p$-state, which
belongs to $S_{\rm I-II}$-boundary, will be named as the marginal $\mu$-state.
It exists in the field interval $\Delta_p=h_1-h_r=0.1982$ simultaneously with
M-state  and is depicted in Fig.7(a) by curve $\mu$ at the field of a jump
($h_r$) to the Meissner state M$_\mu$.

If one moves from $S_{\rm I-II}$-boundary into region I$_b$, a superconducting
$p$-state begins nucleating from a supercooled $n$-state at the field $h_p$.
The amplitude of $p$-state reaches the maximum at the field $h_r$, after that
the jump to M-state occurs. Such hysteresis $p$-state (for $\kappa=0.8$, region
I$_b$) is represented by curve $p_r$ at the point of a jump ($h_r=0.5664$) to
M$_p$-state (not shown); $p$-state with $\kappa=0.8$ exists in the field
interval $\Delta_p=h_p-h_r=0.0744$.

The field interval $\Delta_p$, where $p$-state exists, diminishes rapidly with
$\kappa$ diminishing. (At $\kappa=0.708$, region I$_b$, we have $h_p=0.5018$,
$h_r=0.5013$, $\Delta_p=5\cdot 10^{-4}$; at $\kappa=0.707$, region I$_a$, we
have $\Delta_p=0$, with M-state restoring from a supercooled $n$-state in a
first order jump without forming $p$-state preliminary.)\, The last of
$p$-states existing in region I$_b$ corresponds to $\pi$-boundary of Fig. 1.
Such $\pi$-state is represented by curve $\pi$ in Fig. 7(a) (at the field of a
jump into corresponding M$_\pi$-state, $h_r=0.5013$).

The profiles of $b(x)$ for the same states ($D_\lambda=10$) are shown in
Fig. 7(b).

If $D_\lambda$ diminishes, the interval of $\kappa$ in region I$_b$ (where the
hysteresis $p$-states exist) diminishes also (Fig. 1) and $\pi$-boundary curve
merges with the curves $S_{\rm I-II}$ and $\zeta$ at the same point $G$.
There are no $p$-states in region I$_a$.

It is interesting to watch how the profiles $\psi(x)$ and $b(x)$ of
$\pi$-states change, while moving along $\pi$-boundary in Fig. 1. This is shown
in Fig. 8(a,b) where $\pi$-solutions for $D_\lambda=10$ ($\kappa_\pi=0.708$,
$h_r=0.5013$), $D_\lambda=7$ ($\kappa_\pi=0.708$, $h_r=0.5014$), $D_\lambda=5$
($\kappa_\pi=0.708$, $h_r=0.5015$) and $D\lambda=3$ ($\kappa=0.775$,
$h_r=0.6152$) are presented. [In distinction to Fig. 7(b) where
$b=B/H_\lambda$, the fields $b_\xi=B/H_\xi$ in Fig. 8(b) are normalized to
$H_\xi=\kappa^2H_\lambda$.]

The dotted curve in Fig. 8(a) is the solution ($W$) of the linear equation (9),
normalized to the maximum value of curve {\it 3}. It is evident, that the
self-consistent solutions $\psi(x)$ are described at $\psi_0\ll 1$ by the
Weber functions (10). [Simultaneously, $b_\xi\approx 1$ and $H\approx H_\xi$,
so the linear equation (9) can be used for finding the minimal supercooling
field $H_r(D)$, with $H_r(D)\to H_\xi$ at $D\gg 1$].

One can see from Fig. 8(a), that when $\kappa\to 1/\sqrt{2}$ and
$D_\lambda\gg 1$ the $\pi$-states profiles take a characteristic shape of the
interface between $s$- and $n$- half-spaces [1]. We show below that in this
special case the metastable (hysteresis) $\pi$-state coincides with the
degenerate Bogomolnyi state [8] and can be described analytically [9].

\centerline{}
\centerline{\bf 7. Connection with the Bogomolnyi equations }
\centerline{}

As was shown by Bogomolnyi [8], at $\kappa=1/\sqrt{2}$ the
GL-equations for the infinite superconductor degenerate and can be reduced
to a system of two nonlinear first order differential equations, which have
the analytic solutions [8,9]. If $\psi$ is a real function only of one
Cartesian co-ordinate $x$, the solution is given by the implicit formula [9]:
$$
\int_{\psi_i}^{\psi}{dy\over{y\sqrt{y^2-(1+{\rm ln}\,y^2)}}}=\pm x, \quad
b_\xi^2(x)=1-\psi^2(x),\quad  b_\xi(x)={ B(x) \over H_\xi},       \eqno(11)
$$
with $\psi\to 1$ if $x\to -\infty$, and $\psi\to 0$ if $x\to +\infty$.
Point $x=0$ is defined by the condition $d^2\psi/dx^2$=0 (the inflexion point
of $\psi(x)$), i.e., by the equation $\psi^2-1-{\rm ln}\,\psi=0$ with the
root $\psi_i=0.451$.

The solid line in Fig. 9 is the solution $\psi(x)$ found from the full system
of GL-equations ($\pi$-solution in Fig. 8(a) at $D_\lambda=10$, $\kappa=0.708$).
The dotted line is the analytic solution (11) $(-\infty<x<+\infty$,
$\kappa=1/\sqrt{2}$, the inflexion points of both solutions are superimposed).
Evidently, the self-consistent $\pi$-solution in the limit $D_\lambda\gg 1$
coincides with the Bogomolnyi solution. The self-consistent field $b_\xi(x)$
(see the solution with $D_\lambda=10$ in Fig. 8(b)) also matches the formula
(11) ($b^2_\xi=1-\psi^2$). [In this connection see [17], where the solutions
of the Bogomolnyi equations for a single vortex in the infinite superconductor
are discussed.]

In Fig. 7 the profiles of $p$-solutions $\psi(x)$ and $b(x)$ were shown at the
point of a jump $h_r$ from $p$- to M-state. Fig. 10 illustrates the behavior of
$p$-solutions as function of $h_\xi$ for $D_\lambda=10$ and various $\kappa$.
Figs. 10(a,b) demonstrate: (a) -- the mean value $\overline{\psi}$ and (b) --
the free energy $\Delta g$ (b) in  $\mu$-state ($\kappa_\mu=0.922$, the
peephole is on the boundary $S_{\rm I-II}$). The field interval
$\Delta_p=h_p-h_r$ where $\mu$-state exists is $\Delta_p=0.2331$
(normalized by $H_\xi$). In Figs. 10(c,d) ($\kappa_p=0.8$, the peephole is
inside region I$_b$) this interval is $\Delta_p=0.1163$. In Figs. 10(e,f)
($\kappa_\pi=0.708$, the peephole is almost on $\pi$-boundary) $\Delta_p=0.001$.

From Fig. 10 it follows: 1) the degenerate Bogomolnyi solution (B-state, which
exists at $\kappa=1/\sqrt{2}$, $D_\lambda\to\infty$) is a special case of
$p$-states nucleating in the hysteresis FD-regime from a supercooled $n$-state;
2) B-state exists only in the field $h_1=h_p=h_r=1$ (i.e., $H=H_\xi$) when
the superconductivity simultaneously originates ($\overline{\psi}_p\approx 0$)
and reaches the maximal amplitude ($\psi(0)=1$, but $\overline{\psi}=0.5$, see
Fig. 9); 3) B-state is metastable because M-state ($\overline{\psi}=1$)
of smaller energy exists as well.

Notice, that Fig. 10 indicates also to the nonanalyticity of GL-solutions at
the degeneration point ($\kappa_0=1/\sqrt{2}$, $D=\infty$, $h_\xi=1$). Indeed,
at $\kappa>\kappa_0$\, $p$-solutions exist, having a form very similar to the
degenerate Bogomolnyi solution (with $\overline{\psi}\approx 0.5$, see Fig. 7).
However, at $\kappa<\kappa_0$ only the absolutely stable M-state remains
with $\overline{\psi}\approx 1$, and absolutely unstable $n$-state with
$\psi\equiv 0$. In another words, there is a termination point of $p$-solutions
on $\pi$-boundary, i.e., nonanalyticity at the point $\kappa=\kappa_0$.

\centerline{}
\centerline{\bf 8. Critical fields (phase diagrams)}
\centerline{}

As was mentioned, in an arbitrary point of the state diagram (Fig. 1) the
critical fields exist ($h_1$ and $h_2$ in FI-regime, or $h_p$ and $h_r$ in
FD-regime) which are represented schematically in Fig. 2. The dependence of
the critical fields on the plate thickness may be seen, if one makes a mental
cut of the plane of states in Fig. 1 along a line $\kappa={\rm const}$. The
picture seen is presented in Fig. 11 by a number of phase diagrams (in the
co-ordinates $h_\xi=H/H_\xi$, $D_\xi=D/\xi$) for plates with different $\kappa$.

Dashed line $\kappa=0.3$ in Fig. 11(a) corresponds to the critical field $h_1$
(FI-regime); below this line lies the region of superconducting M-phase
($\overline{\psi}\sim 1$), above this line lies the region of $n$-phase
($\overline{\psi}\equiv 0$). Solid line $W$ corresponds to the critical field
$h_r$ (FD-regime); above this line lies the region of metastable $n$-phase,
below this line lies the region of M-states. In region I$_a$ (Fig. 1) a
supercooled $n$-state becomes absolutely unstable in the field $h_r$ where the
jump to M-state occurs. A supercooled $n$-state exists in the field interval
$\Delta_n=h_1-h_r$ where the hysteresis is possible. The interval $\Delta_n$
diminishes with the plate thickness and at some $D_\xi$ the lines $h_1$ and
$h_r$ merge ($\Delta_n=0$). This point in Fig. 1 (at $\kappa=0.3$) is
represented by the value $D_\lambda=1.14$ ($D_\xi=\kappa D_\lambda=0.342$),
which lies on $\zeta$-boundary of the hysteresis region. For smaller
thicknesses (region II$_b$ in Fig. 1) there exists unique critical field $W$,
in which the superconducting state is destroyed (FI) or originates (FD)
without hysteresis by a second order phase transition. (Notice, that for small
$\kappa$ the hysteresis interval $\Delta_n$ seems to increase, because the
field scale diminishes, $H_\xi=\kappa^2H_\lambda$.)

In Fig. 11(a) the critical fields $h_1$ and $W$ are also drawn for the values
$\kappa=0.5$ and $\kappa=0.7071$ (which lie to the left of the line
$\kappa_0=1/\sqrt{2}$ in Fig. 1). Evidently, the critical field $h_1$ (dashed
line) diminishes with $\kappa$ increasing. The fields $h_r$ (normalized to
$H_\xi$) are represented for all $\kappa$ [12] by a single curve $W$ (which
corresponfs to the stability boundary of a supercooled $n$-phase).

The picture changes in passing to $\kappa>\kappa_0$. Because in region I$_b$
(Fig. 1) there are metastable $p$-states, we have three critical fields here
($h_1,h_p,h_r$ in Figs. 2(b,f)). These fields are represented in Fig. 11(a)
(for $\kappa=0.8$) by three curves $h_1,W,h_r$. The field $h_1$ (dashed line)
corresponds again to the maximum field, at which the jump from M- to $n$-state
occurs. At the field $h_p$ a supercooled $n$-state becomes absolutely unstable
and a superconducing (metastable) $p$-state of small amplitude originates
(this field is presented by the same curve $W$ as for $\kappa<\kappa_0$). At
the field $h_r$ (dotted line) the metastable $p$-state becomes absolutely
unstable and a first order transition from $p$- to M-state occurs. In the field
interval $\Delta_p=h_p-h_r$ the metastable $p$-states exist and the hysteresis
is possible.

The field interval $\Delta_p$ diminishes with the plate thickness, so crossing
$\pi$-boundary (Fig. 1) we get from region I$_b$ into region I$_a$ where
$p$-states are absent but the supercooled $n$-state exists. (For $\kappa=0.8$
$\pi$-boundary is crossed at $D_\lambda=2.7$ or $D_\xi=2.16$. The dotted line
$h_r$ in Fig. 11(a) merges with the curve $W$ at point $\alpha$.)\, In region
I$_a$ there are only two critical fields ($h_1$ and $W$). Further decreasing
$D$ we get from region I$_a$ into region II$_b$ crossing the hysteresis
$\zeta$-boundary (for $\kappa=0.8$ this happens at $D_\xi=1.2$, point $\beta$
in Fig. 11(a), when the curves $h_1$ and $W$ merge). In region II$_b$
($D_\xi<1.2$) no hysteresis is possible and in Fig. 11(a) remains only one
critical curve $W$ which describes the states with $\psi\to 0$.

In Fig. 11(b) the critical fields are depicted for $\kappa=1;\, 1.2;\, 2$.
If $\kappa=1$ (the mental cut in the plane of Fig. 1 lies inside region II$_a$)
there are three critical fields: $h_r, h_1, h_2$ (see Fig. 2(c,g)). The tail of
$e$-states with $\overline{\psi}\ll 1$ either vanish (FI) or appears (FD) in
the field $h_2$. This field is marked in Fig. 11(b) by a letter $W$
(solid curve). In the field $h_1$ (dashed line) the jump from M- to $e$-state
occurs (in FI-regime) and $d$-state appears (in FD-regime). The metastable
$d$-states (and the corresponding hysteresis) exist down to the field $h_r$
where the jump from $d$- to M-state occurs. The width of the hysteresis region
$\Delta_d=h_1-h_r$ depends on the plate thickness. The points $\Delta_d=0$ in
Fig. 11(b) (or the Landau points, $L$) correspond to the intersection of
the line $\kappa={\rm const}$ with $\zeta$-boundary in Fig.1. (For $\kappa=1$
$\Delta_d=0$ at $D_\xi=2.0$.)\, In region II$_b$ the hysteresis is absent, so
here exists only one critical field $W$ which describes the reversible
second order phase transition.

Phase diagrams at $\kappa=1.2$ and $\kappa=2$ are analogous to the case
$\kappa=1$. The second order phase transition curve $W$ (as well as the curve
$W$ in Fig. 11(a)) is the same for all $\kappa$, it can be found from the
linearized equation (9) and expressed through the Weber functions $w$ (10).
However, to find the first order phase transition fields ($h_1$ and
$h_r$) it is necessary to solve full system of GL-equations.

Note, that the interval between $L$-point (where $\Delta_d=0$) and $W$-curve
diminishes with $\kappa$, so the curves of Fig. 11(b) transform continously
into the curves of Fig. 11(a).

\centerline{}
\centerline{\bf 9. Conclusion}
\centerline{}

Note in conclusion, that the vortex-free states, studied above, may be
realized in mesoscopic samples with characteristic dimension $D$ of several
$\lambda$. With $D$ increasing the uniform (one-dimensional) edge-suppressed
$e$-state (as well as $d$- and $p$-states) may become unstable relative
breaking the boundary region into separate vortices, with forming subsequently
the regular vortex lattice [2]. However, the detailed study of such
inhomogeneous states demands the solution of partial differential
equations what is outside the scope of the present investigation. (In this
connection see, for instance, Refs. [18,19] where some of such problems are
considered to explain the experiments [20--24] with thin superconducting discs
of various form in a perpendicular magnetic field.)\, Besides, many of the
theoretical results happen to depend rather weakly on the sample geometry, so
the predictions obtained for the plate (or the cylinder [3]) on the base of
one-dimensional equations, have, probably, more general value and may be used
in discussing the details of the concrete experiments.

I am gratefull to V. L. Ginzburg for the interest in this work and valuable
comments, and also to A.Yu. Tsvetkov and V. G. Zharkov for discussions. This
work was supported through the grant RFFI 02-02-16285.

\vfill\eject

\centerline{}
\centerline{\bf References}
\centerline{}

[1] V.L.Ginzburg, L.D.Landau, Zh.Exp.Teor.Fiz. {\bf 10}, 1064 (1950).

[2] A.A.Abrikosov, {\it Fundamentals of the Theory of Metals}
(North-Holland, Amsterdam,

\quad 1988).

[3] G.F.Zharkov, Zh.Exp.Teor.Fiz. {\bf 122}, $N$9 (2002).

[4] P.M.Markus, Rev. of Mod. Phys., {\bf 36}, 294 (1964).

[5] A.Yu.Tsvetkov, G.F.Zharkov, V.G.Zharkov, Krat.Soob.Fiz. FIAN, $N$2, 42 (2002).

[6] G.F.Zharkov, V.G.Zharkov, A.Yu.Zvetkov, cond-mat/0008217 (2000);

\quad Phys.Rev.B, {\bf 61} 12293 (2000).

[7] G.F.Zharkov, JLTP, {\bf 128}, $N$3/4 (2002).

[8] E.B.Bogomolnyi, Yad.Fiz. {\bf 24}, 861 (1976)
[Sov.J.Nucl.Phys. {\bf 24}, 449 (1976)];

\quad E.B.Bogomolnyi and A.I.Vainstein, {\it ibid.} {\bf 23}, 1111 (1976)
[{\bf 23}, 588 (1976)].

[9] I.Luk'yanchuk, Phys.Rev.B, {\bf 63}, 174504 (2001).

[10] G.F.Zharkov, V.G.Zharkov, Physica Scripta {\bf 57}, 664 (1998).

[11] E.Kamke, "{\it Differentialgleichungen}", part I, Leipzig (1959).


[12] D.Saint-James, P.deGennes, Phys.Lett. {\bf 7}, 306 (1963).

[13] D.Saint-James, Phys.Lett. {\bf 15}, 13 (1965).

[14] V.L.Ginzburg, Zh.Exp.Teor.Fiz, {\bf 34}, 113 (1958) [Sov.Phys.JETP,
{\bf 34}, 78 (1958)].

[15] L.D.Landau, Phys.Zs.Sowjetunion {\bf 8}, 113 (1935); {\bf 11}, 26 (1937).

[16] H.J.Fink, D.S.McLachlan and B.Rothberg-Bibby, in {\it Prog. in
Low Temp.Phys.},

\quad v. VIIb, p.435, ed. D.Brewer, North Holland Pub., Amsterdam- N.Y.- Oxford (1978).

[17] V.Hakim, A.Lema\^{\i}tre, K.Mallick, Phys.Rev.B, {\bf 64}, 134512 (2001).

[18] V.A.Schweigert, F.M.Peeters et.al., Phys.Rev.Lett. {\bf 79},
4653 (1997);

\quad ibid, {\bf 83},2409 (1999); Supralatt. and
Microstruct. {\bf 25}, 1195 (1999);

\quad Phys.Rev.B {\bf 59}, 6039 (1999); ibid, {\bf 62}, 9663 (2000);
Physica C, {\bf 332}, 266,426,255 (2000).

[19] J.J.Palacios, Phys.Rev.B {\bf 57}, 10 873 (1998); Physica B,
{\bf 256-258}, 610 (1998);

\quad Phys.Rev.Lett. {\bf 83}, 2409 (1999); {\bf 84},1796 (2000).

[20] V.V.Moshchalkov et al., Nature (London) {\bf 373}, 319 (1995);
{\bf 408}, 833 (2000).

[21] A.K.Geim et al., Nature (London) {\bf 390}, 259 (1997);
{\bf 396}, 144 (1998); {\bf 407}, 55 (2000);

\quad Phys.Rev.Lett., {\bf 85}, 1528 (2000); {\bf 86}, 1663 (2001).

[22] D.S.McLachlan, Solid State Commun., {\bf 8}, 1589, 1595 (1970).

[23] O.Buisson et al., Phys.Lett.A {\bf 150}, 36 (1990).

[24] F.B.M\H{u}ller-Allinger, A.C.Motta, Phys.Rev.B {\bf 59}, 8887 (1999).
\vfill\eject

\centerline{}
\centerline{\bf Figures captions}
\centerline{}

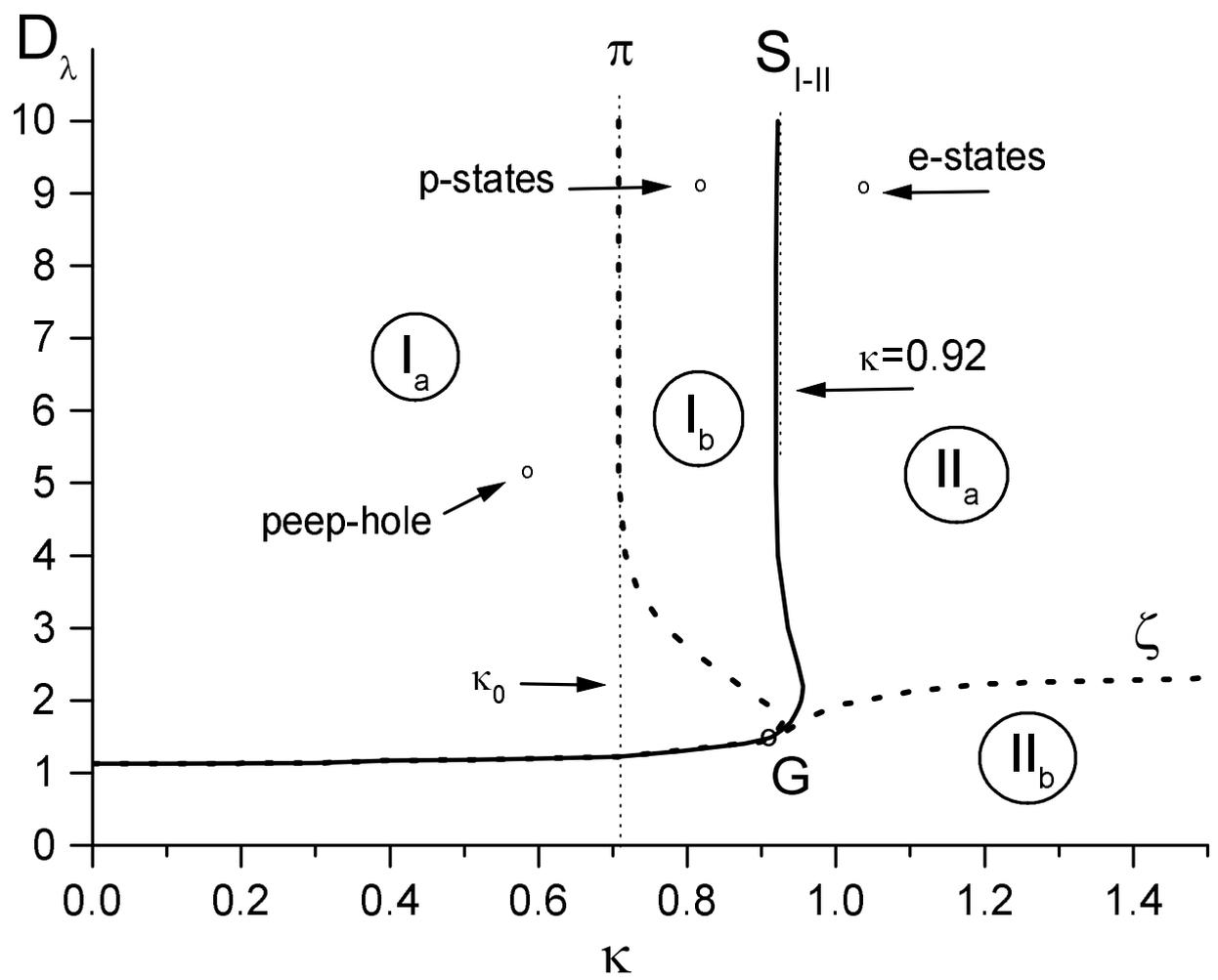
Fig. 1. The state diagram on the plane ($D_\lambda, \kappa$). Curve
$S_{\rm I-II}$ is the boundary between first and second order phase transitions
from $s$- to $n$-state in the increasing field (FI-regime); curve $\pi$ is the
boundary of metastable $p$-states in the decreasing field (FD-regime); curve
$\zeta$ is the hysteresis boundary, below $\zeta$-line the hysteresis is absent.
The asymptotics of $\zeta$-boundary are: $D_\lambda=1.13$ for $\kappa\to 0$;
$D_\lambda=2.43$ for $\kappa>6$.

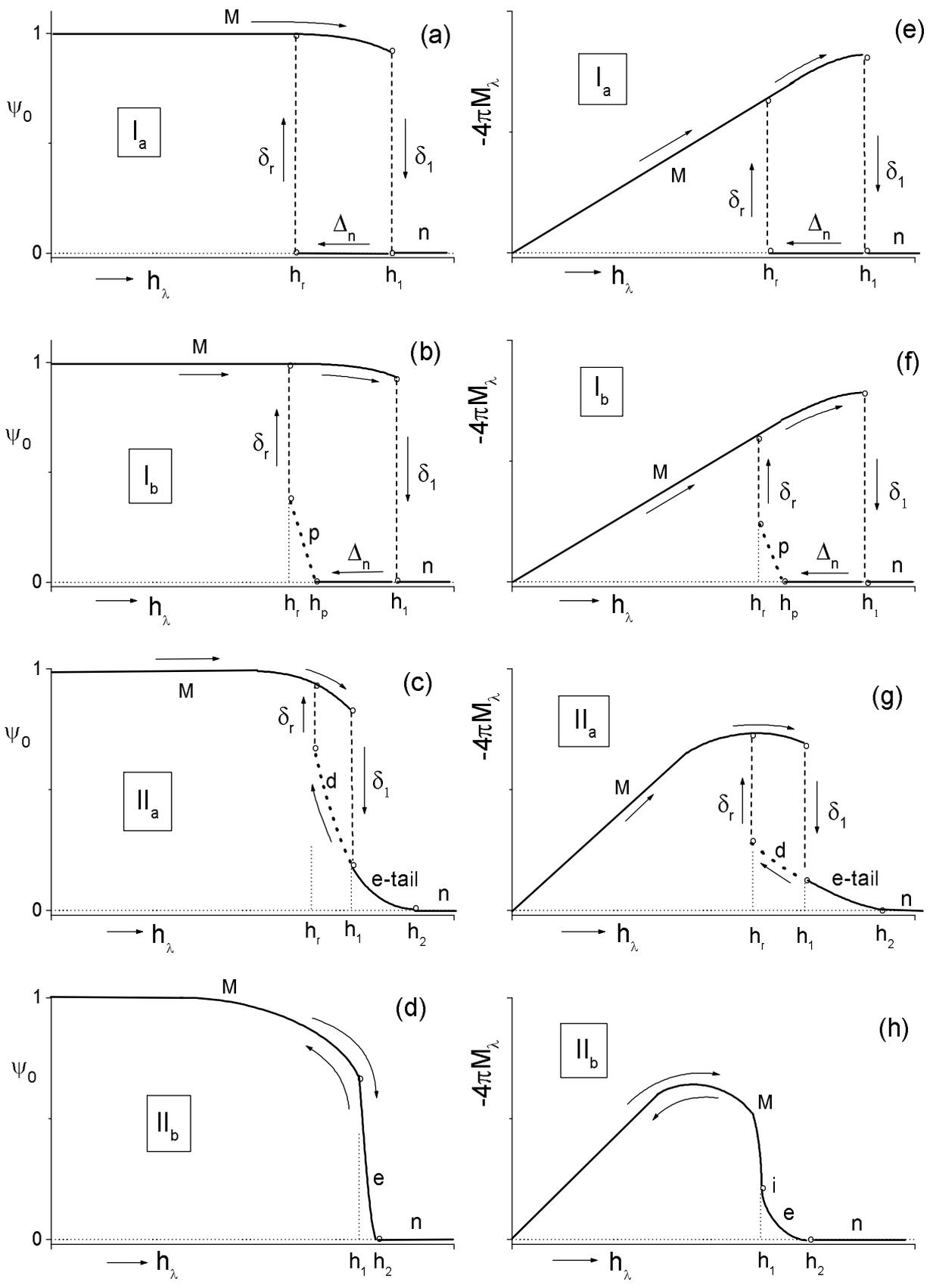
Fig. 2. The order parameter $\psi_0$ and magnetization $(-4\pi M_\lambda)$
versus field $h_\lambda$ in different regions of Fig. 1 (schematically, see the
text).

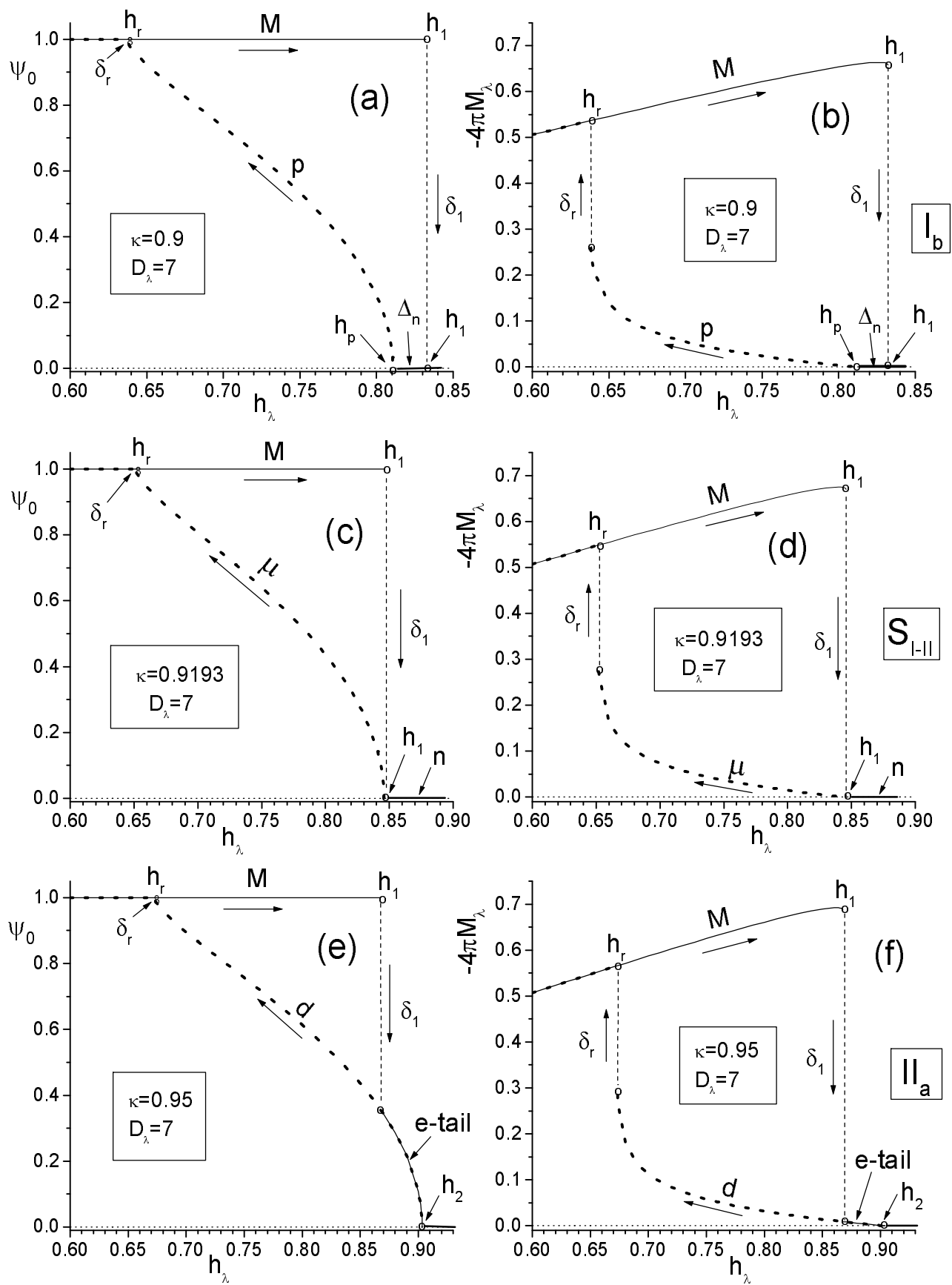
Fig. 3. The dependences of $\psi_0$ and $-4\pi M_\lambda$ on $h_\lambda$ in the
vicinity of $S_{\rm I-II}$-boundary in Fig. 1 ($D_\lambda=7$, the values of
$\kappa$ are given in the figure.) M is the Meissner state; $p$ is the precursor
state (region I$_b$); $\mu$ is the marginal $p$-state (laying on the critical
$S_{\rm I-II}$-boundary); $d$ is the metastable depressed $d$-state (region
II$_a$). Dotted lines correspond to FD-regime.

¨á. 4. The dependences of $\psi_0$ and $-4\pi M_\lambda$ on $h_\lambda$ at
$D_\lambda=2$ and various $\kappa$: $\kappa=0.93$ (region I$_b$ in Fig. 1),
$\kappa=0.953$ ($\mu$-state on $S_{\rm I-II}$-boundary), $\kappa=0.98$ (region
II$_a$), $\kappa=1.03$ ($\zeta$-boundary), $\kappa=1.05$ (region II$_b$);
$p$, $\mu$, $d$ are the hysteresis (metastable) states. On the curves
$\kappa=1.05$ the points of inflexion $i$ are marked.

¨á. 5.\, (a) -- The dependences $\psi_0(h_\lambda)$ and
$-4\pi M_\lambda(h_\lambda)$ in $G$-point in Fig. 1.\, (b) -- The dependences
$\psi_0(h_\lambda)$ in the vicinity of $G$-point ($D_G=1.51$): {\it 1} --
$\kappa=0.8$ (region I$_a$); {\it 2} -- $\kappa=0.915$ ($G$-point);
{\it 3} -- $\kappa=1.0$ (region II$_b$).

¨á. 6. (a) -- The order parameter $\psi(x)$ and (b) -- the field
$b_\lambda(x)$ ($D_\lambda=7$) in different states:
$e$ -- $\kappa=1.1$, $h_1=0.9743$;
$p$ -- $\kappa=0.8$, $h_r=0.5676$; $\mu$ -- $\kappa=0.9193$, $h_r=0.6520$;
$d$ -- $\kappa=1.1$, $h_r=0.7796$. The corresponding M-states (see the text)
are also shown.

(c) and (d) -- The analogous curves for $D_\lambda=2$: $e$ -- $\kappa=0.97$,
$h_1=1.0172$; $p$ -- $\kappa=0.93$, $h_r=0.9612$; $\mu$ -- $\kappa=0.953$,
$h_r=0.9903$; $d$ -- $\kappa=0.97$, $h_r=1.072$; $i_0$ -- $\kappa=1.030$,
$h_\lambda=1.076$; $i$ -- $\kappa=1.05$, $h_\lambda=1.1$ (see the text).

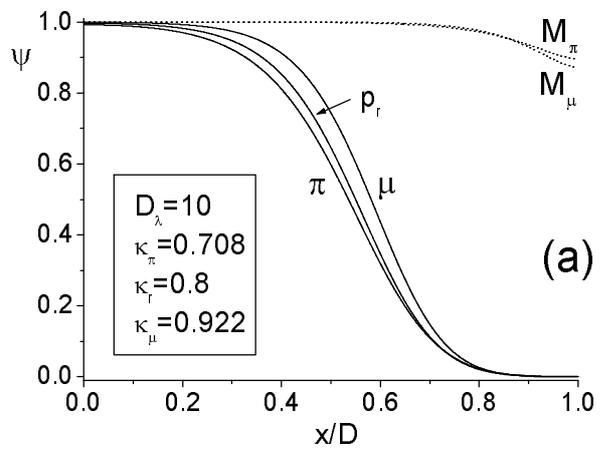
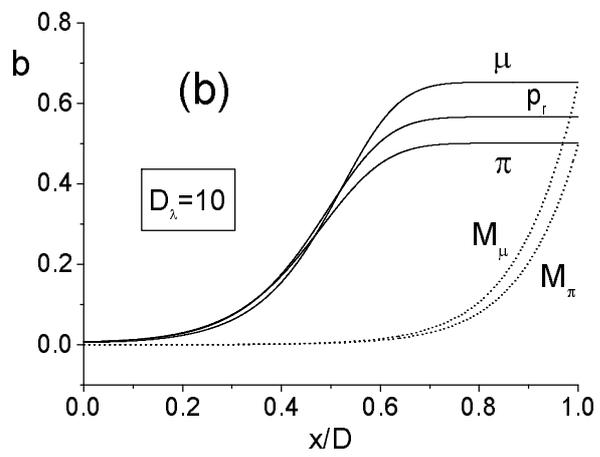
Fig. 7. The precursor states: (a) -- $\psi(x)$ and (b) -- $b_\lambda(x)$ for
$D_\lambda=10$ and different $\kappa$: $\mu$ -- ($S_{\rm I-II}$-boundary)
$\kappa=0.922$, $h_r=0.6529$; $p_r$ -- $\kappa=0.8$, $h_r=0.5664$; $\pi$ --
$\kappa=0.708$, $h_r=0.5013$ (see the text).

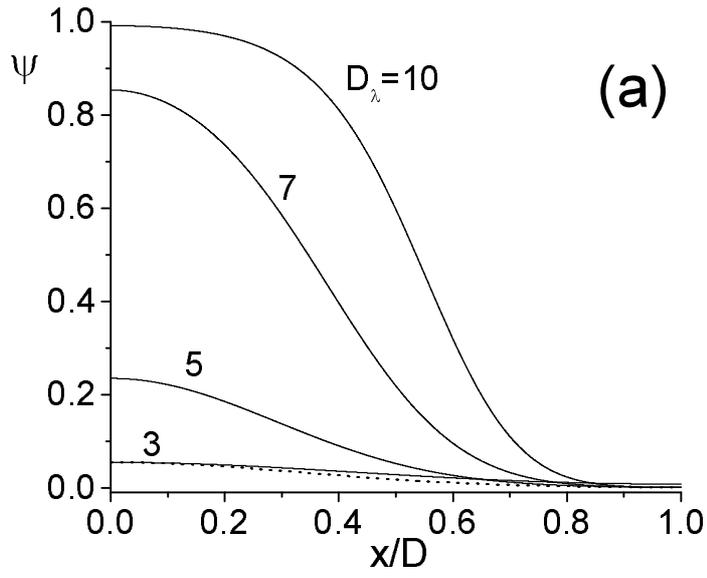
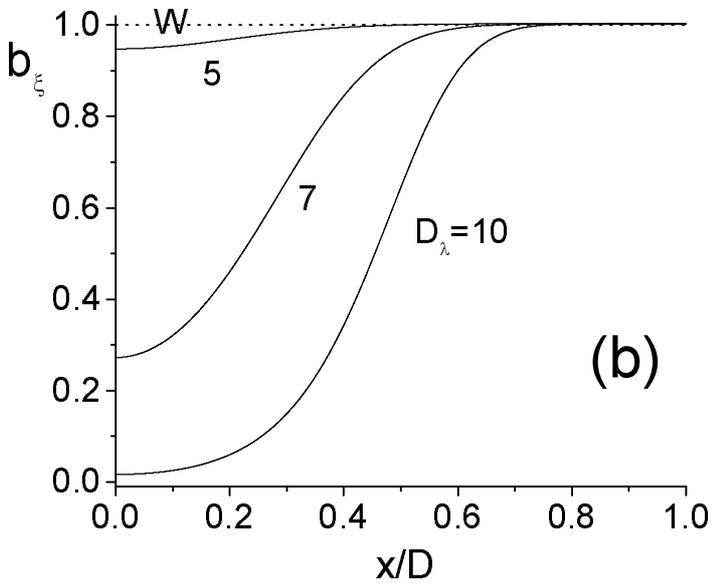
Fig. 8. The space profiles of $p$-states, laying on $\pi$-boundary at the field
of a jump ($h_r$) from $p$- to M-state (for $D_\lambda=5;7;10$). The values of
$\kappa$ and $h_r$ (normalized to $H_\lambda$) are given in the text. The
dotted line in (a) is the normalized solution $W$ of the linear equation (9).

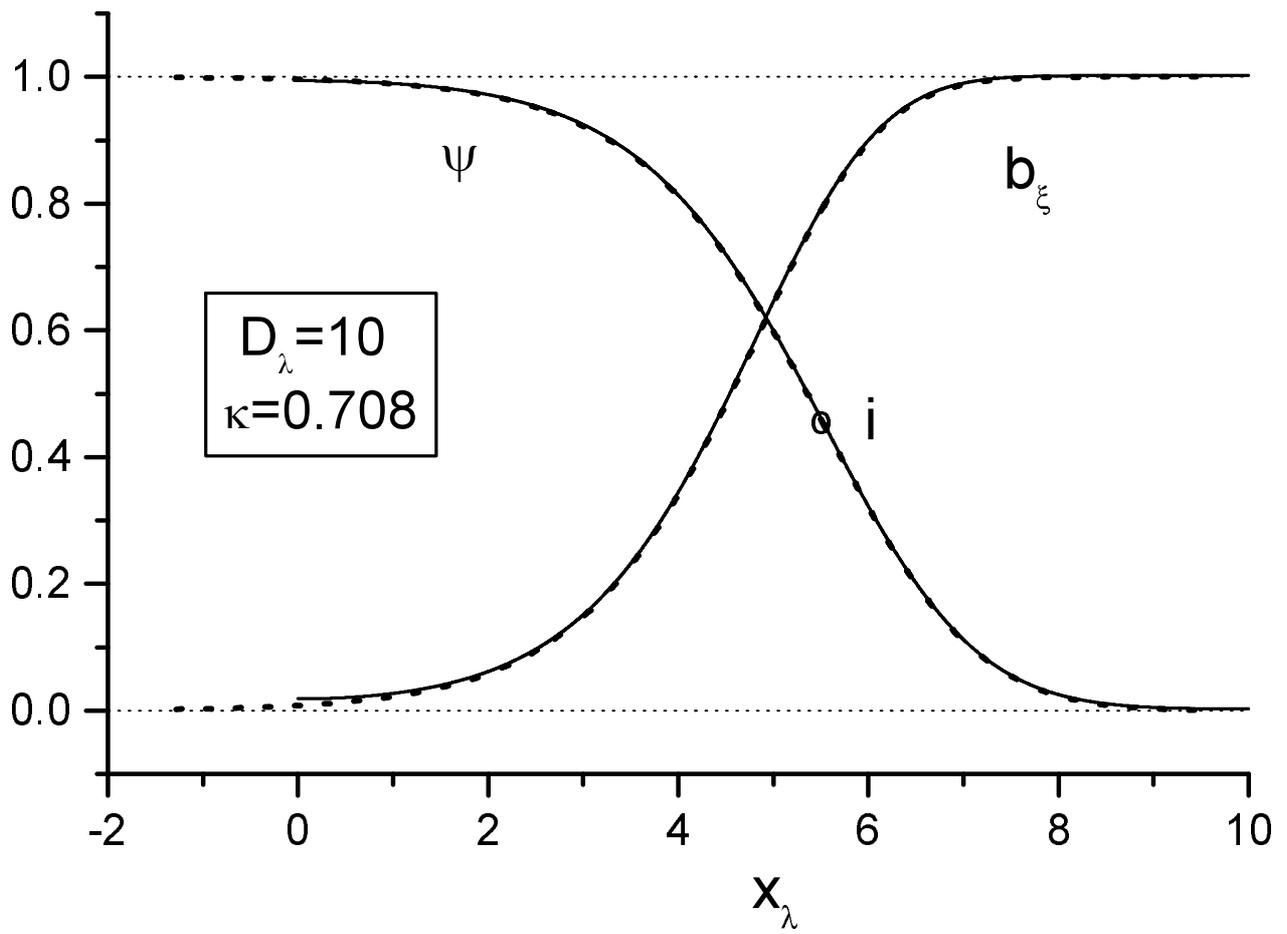
Fig. 9. Solid lines are the self-consistent solutions for $\psi(x)$ and $b(x)$
(normalized to $H_\xi$) for $D_\lambda=10$, $\kappa=0.708$. Dashed line is the
degenerate Bogomolnyi solution (11). The solutions are superimposed at the
inflexion points $\psi_i=0.451$.

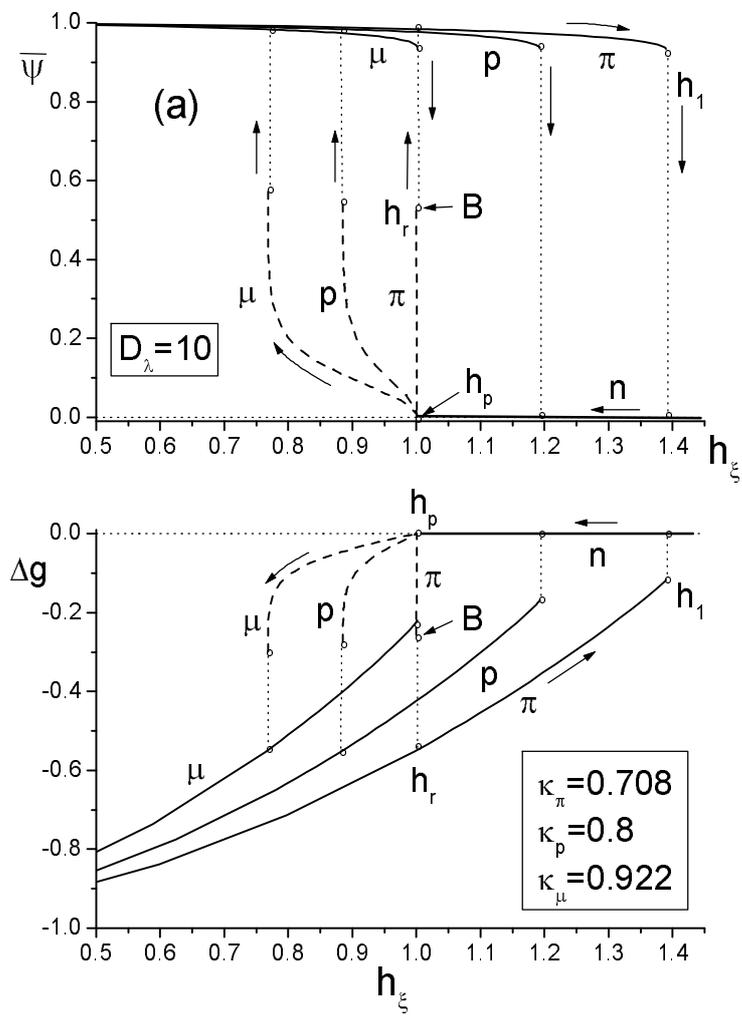
Fig. 10. The mean value $\overline{\psi}$ (a) and the free energy $\Delta g$
(b) as functions of $h_\xi$ for the states existing in the plate of thickness
$D_\lambda=10$ and various $\kappa$ (shown in figure). Solid lines are Œ- and
$n$-states, dashed lines are the precursor states ($\mu, p, \pi$). Points B
on $\pi$-curves correspond to the degenerate Bogomolnyi solution with
$\overline{\psi}\approx 0.5$.

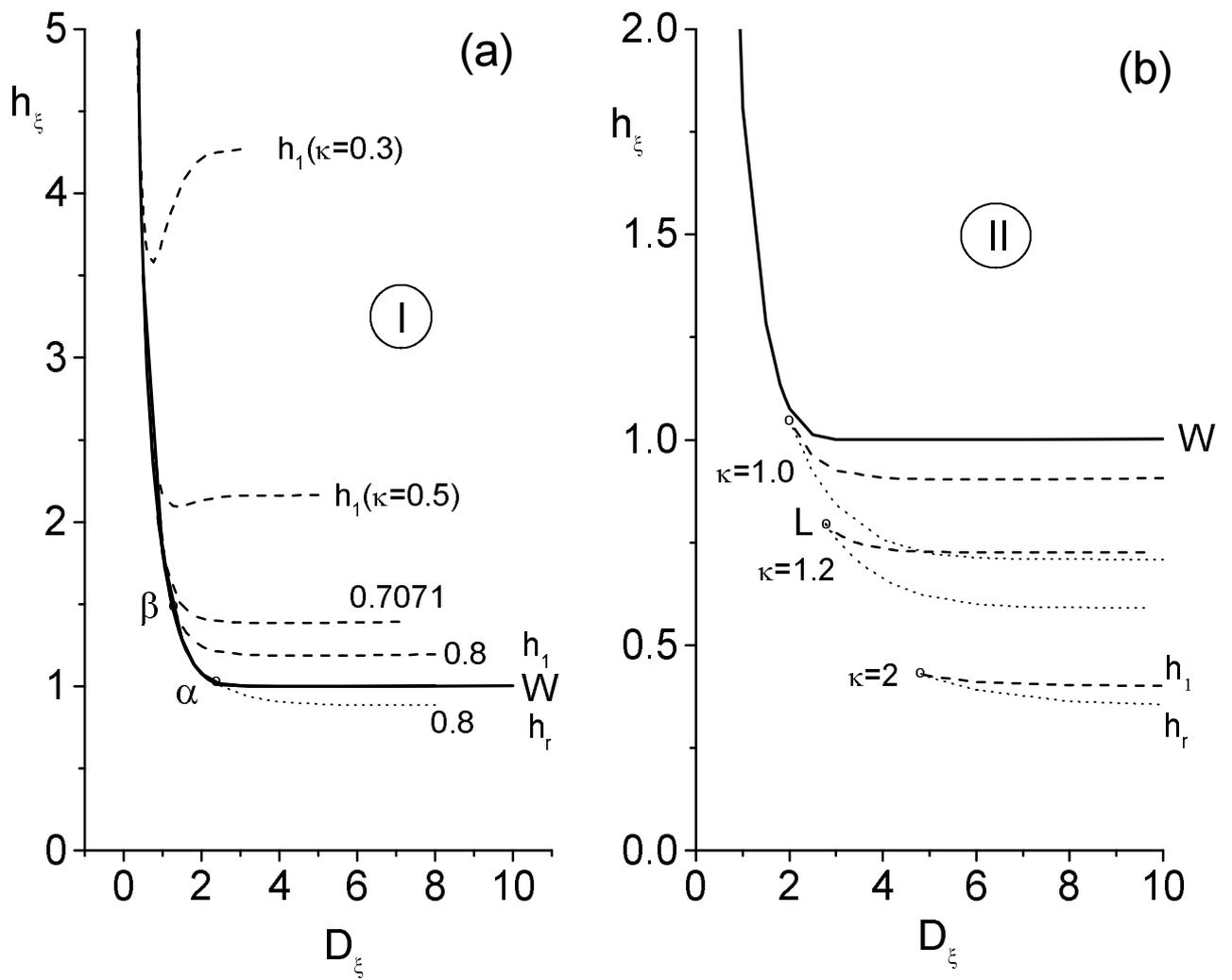
Fig. 11.  The critical fields for different $\kappa$ (shown in figure): (a) --
type-I superconductors, $\kappa<\kappa_c$; (b) -- type-II superconductors,
$\kappa>\kappa_c$. Dashed lines are the fields $h_1$ in which M-state becomes
absolutely unstable (FI); solid line $W$ is the field in which $n$-state
becomes absolutely unstable (FD); dotted lines are the fields $h_r$ in which
M-state restores (FD).

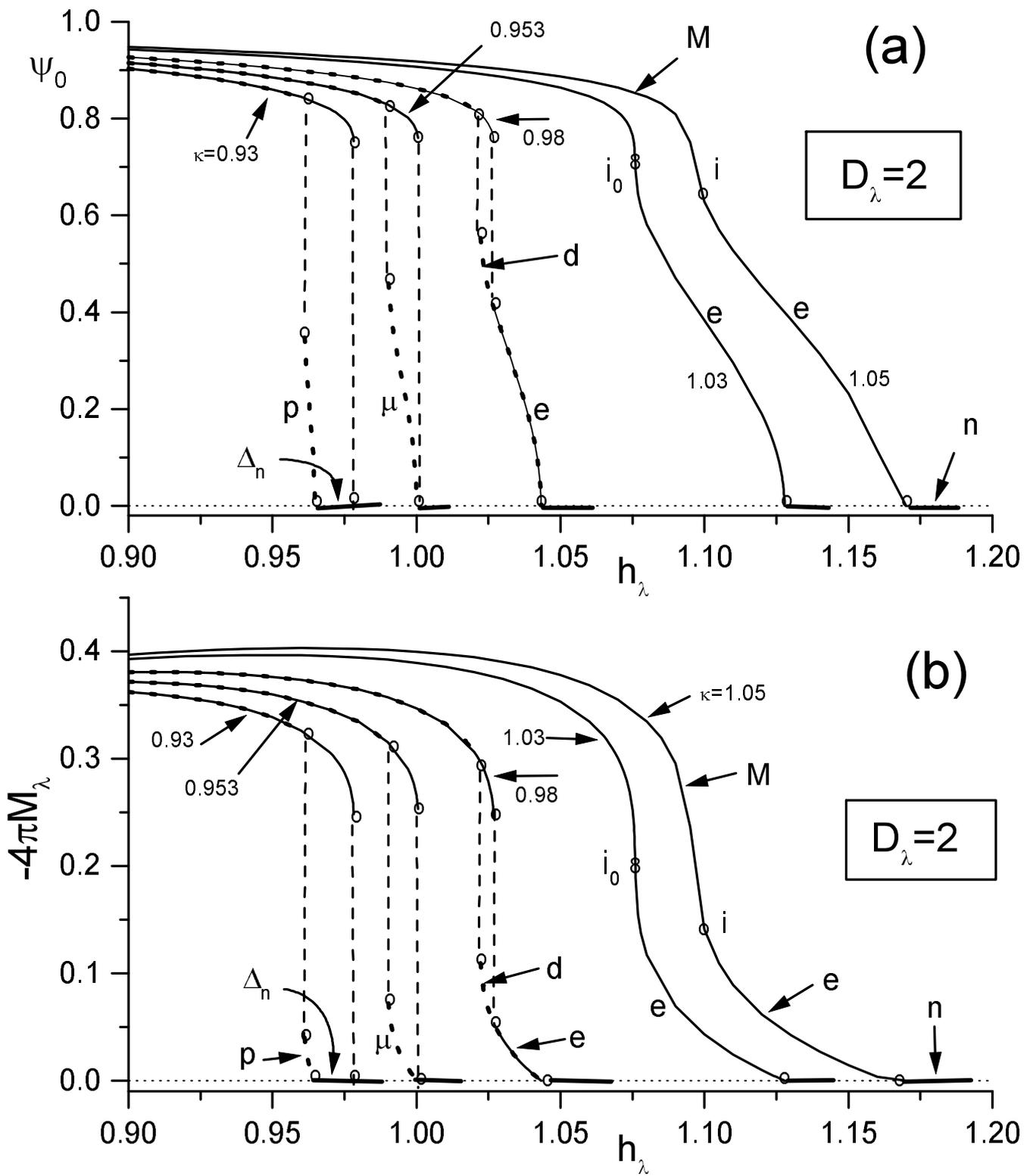
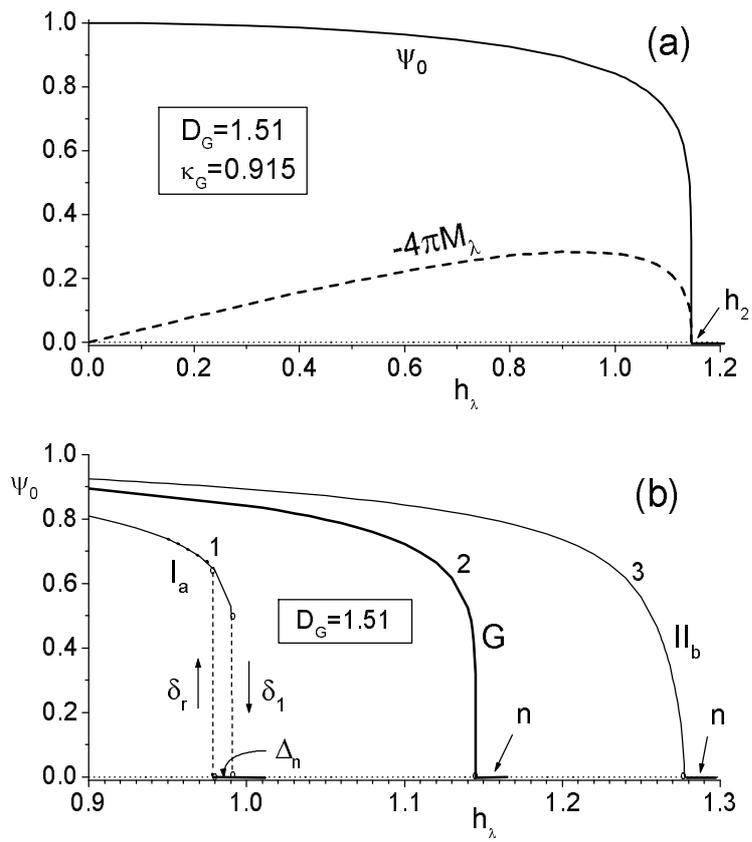
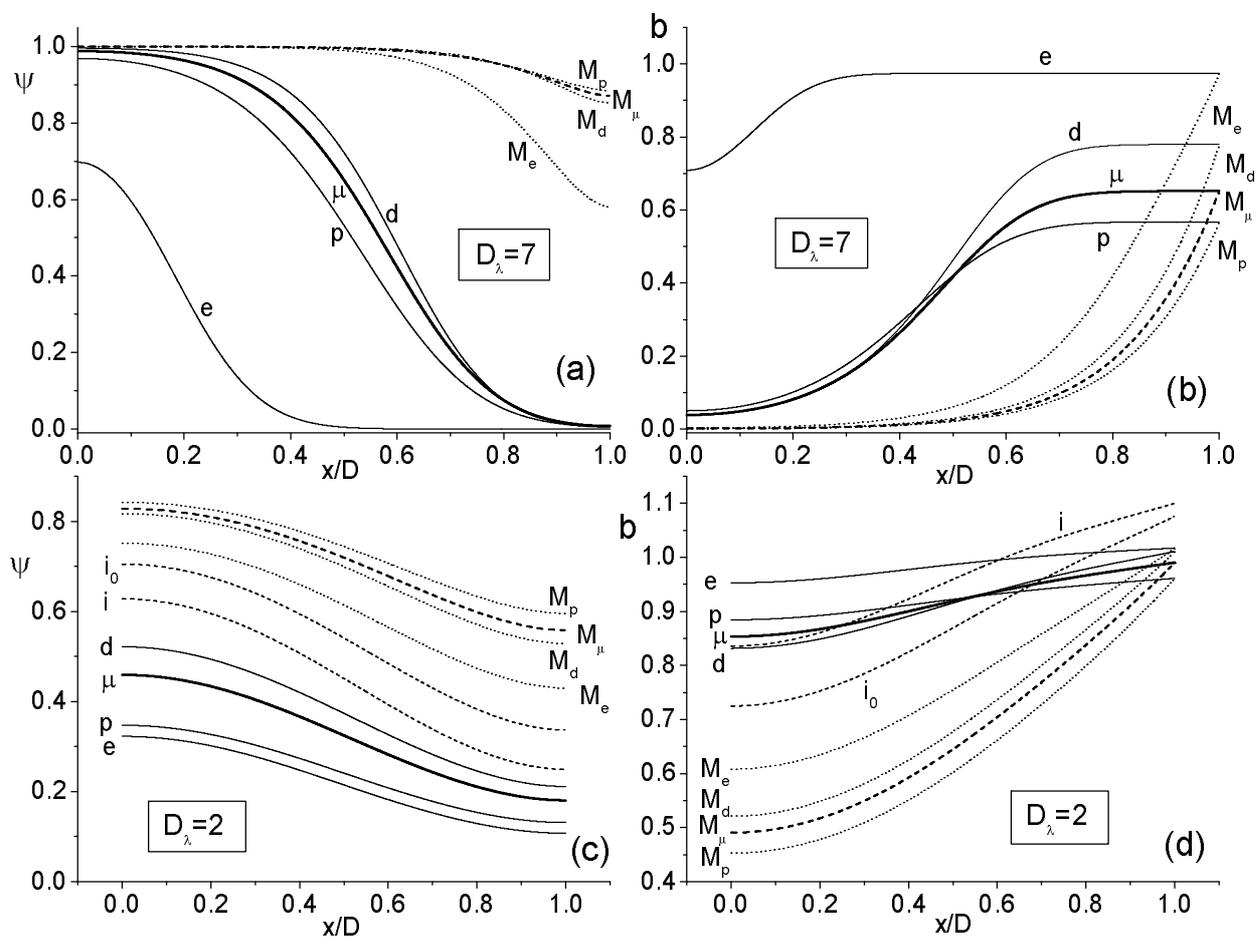
\end{document}